\documentclass{ismdproc}

\def\pythia{{\sc Pythia}}

\def\CASCADE{{\sc Cascade}}
\def\PYTHIA{{\sc Pythia}}
\begin{document}
%------------------------------------
\title{    Introduction to Forward Physics and Cosmic Rays \\ 
\hspace*{6 cm }    at the \\ 
Symposium on Multiparticle Dynamics ISMD 2010}

\author{{\slshape M.~Grothe$^a$, F.~Hautmann$^b$ and S.~Ostapchenko$^c$}  \\[1ex]
$^a$ University of Wisconsin, Madison  \\[1ex]
$^b$ University of Oxford   \\[1ex]
$^c$ NTNU and Moscow State University  }

% please do not modify the following 5 lines
\contribID{xy}  % will be entered by the editors
\confID{yz}
\acronym{ISMD2010}
\doi            % will be entered by the editors

\maketitle

\begin{abstract}
We give a  brief  introduction to the  topics 
discussed at  the ISMD 2010 Symposium 
(Antwerp,  2010) 
on forward physics at the LHC and its  interplay with cosmic rays  physics. 
\end{abstract}

\section{Introduction}

\begin{figure}[htb]
\vspace{20mm}
\includegraphics{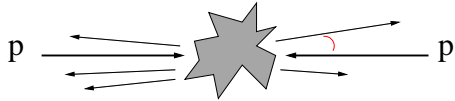}  
\caption{\it Particle production at  forward rapidities in hadronic collisions. } 
\label{fig:forwdef} 
\end{figure}

  Particle production  in the forward region   
  at hadron colliders   (Fig.~\ref{fig:forwdef})     
is traditionally dominated by    low-p$_{\rm{T}}$   physics. 
 At the Large Hadron Collider (LHC), 
due to   the large    center-of-mass energy,  
the  phase space opens  
 up for  high-p$_{\rm{T}}$   forward production.   
 By  exploiting the  unprecedented reach in rapidity 
 of  the experimental instrumentation,  
it    becomes possible,  for the first time  at hadron-hadron colliders, 
 to carry out a program of  
 high-p$_{\rm{T}}$    physics  and jets   in the forward 
 region~\cite{grothe,denterria,ajaltouni},   involving  
 both  new particle discovery 
processes (e.g.,    Higgs 
searches   for    vector boson fusion channels,  jet studies in  decays of    
highly boosted heavy states) and  new aspects of 
standard model physics  (e.g.,  QCD at small x,     
searches for new states of strongly interacting matter at high density).

\begin{figure}[htb]
\vspace{55mm}
\includegraphics{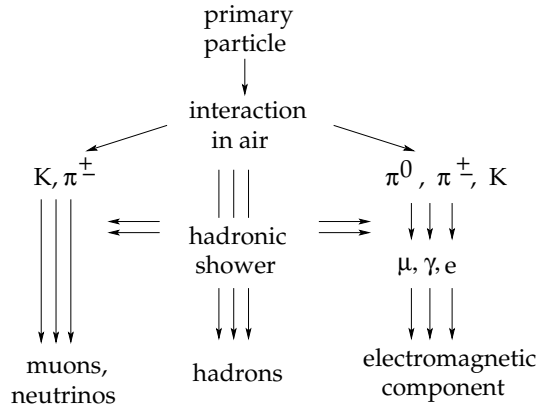}  
\caption{\it High-energy air showers. } 
\label{fig:airsho} 
\end{figure}

Measurements of  forward   particle production   (both soft and hard) 
 at the LHC  are  expected to provide   input  to 
Monte Carlo models  of high-energy air showers~\cite{engel}   for experiments on 
cosmic rays  (Fig.~\ref{fig:airsho}),  as    
LHC pp interactions  correspond  to fixed-target  collisions in air  
in  the midst of the  measured cosmic ray spectrum.

LHC forward physics poses new challenges to 
 both experiment and theory.    On one hand,   
  measurements of   final 
states  boosted to   forward  rapidities   call for new  experimental tools  and techniques.     
 On the other hand, the evaluation of QCD theoretical predictions 
 is made    complex  by  the fact that  the  forward  kinematics 
forces    high-p$_{\rm{T}}$    production  into a 
  region  characterized  by  multiple energy scales,  
possibly widely    disparate from each other.

In this article we start in Sec.~\ref{sec:th}  with a concise overview of  general 
issues  in the  QCD  treatment  of forward hard processes.    In Sec.~\ref{sec:lhc}    
we discuss  a selection  of the  first  measurements   in the forward region  at 
the  LHC presented   at this   meeting.   In  Sec.~\ref{sec:cr}     we address 
  connections of collider measurements   with  cosmic ray physics.  
In  Sec.~\ref{sec:further}  we   describe     further    collider    
studies    discussed at the meeting and  give   final remarks. 

\section{General issues}
\label{sec:th}

  This section  presents briefly 
QCD theory issues  in  the description    of   hard processes in  
the forward region,  introducing  the role of perturbative 
resummations,  of corrections beyond single parton scattering,  and  
  of  methods    that aim  to  extend  the theory towards  
    infrared-sensitive regions.  
See  discussion  in~\cite{rdfield-talk}  for   an introduction to  
      low-p$_{\rm{T}}$   phenomena.

\begin{figure}[htb]
\vspace{45mm}
\includegraphics{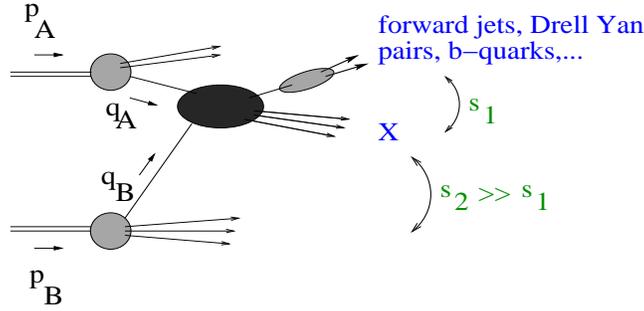}
\caption{\it  Forward   hard production processes 
in hadron-hadron collisions.} 
\label{fig:forwpicture}
\end{figure}

Fig.~\ref{fig:forwpicture}  pictures a      forward hard  event  
in which a      forward jet  (or     some other   high-p$_{\rm{T}}$  probe,   
such as $b$-quark  jets  or Drell Yan pairs) 
is produced    in association with   hard final state $X$. 
See    e.g.  contributions~\cite{sunar,anderson,lykasov,deak}  for 
 specific examples of such events.        
 The kinematics   of the process  in  Fig.~\ref{fig:forwpicture}    is  characterized 
by the  large  ratio  of sub-energies  $s_2  /  s_1 \gg 1 $   
 and  highly asymmetric longitudinal momenta in the partonic initial 
  state   ($x_A \to 1$, $x_B \to 0$).

  In this multiple-scale region,  one       is probing  
 the partonic phase space  near  its  boundaries, and a first,    
basic   question   in  the  QCD treatment of   these  processes   is      
 whether fixed-order perturbative   calculations   accurately describe  
 QCD  theoretical 
 predictions,  or significant contributions arise beyond fixed order which call 
 for perturbative  QCD resummations. 

In the case of jets    at 
forward  rapidity,  it has long been recognized~\cite{muenav}  
that  reliable theoretical  predictions, unlike the case 
of inclusive jet production   in the central region,  require  
  the  resummation of  logarithmically enhanced QCD  corrections, becoming large 
  for asymptotically   high energies.    
  This  early observation  has given rise to an  ample literature  of calculations  
  based on the  use~\cite{muenav} of the 
  BFKL equation.   See~\cite{wallon-etal-10}  for  results 
  at 
  next-to-leading-logarithmic order.  

\begin{figure}[htb]
\vspace{55mm}
\includegraphics{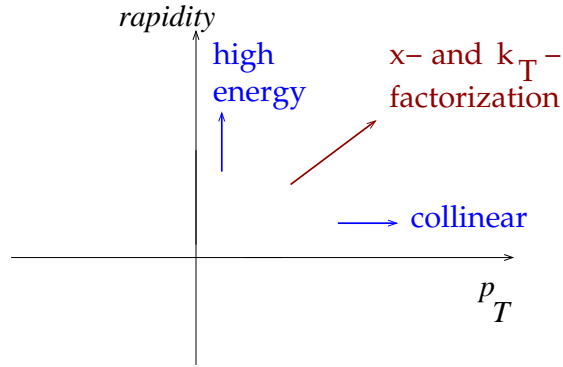}
\caption{\it  QCD  radiative contributions  to forward  hard processes     in  
the rapidity and  transverse momentum plane.} 
\label{fig:ypt}  
\end{figure}

On the other hand,  in forward hard production   at collider energies  
both logarithmic corrections   in  the large  rapidity  interval  
(of  BFKL,  or  high-energy,   type)  
and logarithmic corrections  in  
the hard transverse momentum (of collinear type)
are phenomenologically important~\cite{ajaltouni}.     A pictorial 
representation of  these  radiative contributions  in the 
rapidity and  transverse momentum plane   
is sketched   in  Fig.~\ref{fig:ypt}.  
 The theoretical framework 
to  sum   consistently 
both  kinds of logarithmic corrections   to all  perturbative orders    
   is based on QCD  high-energy    factorization   
 with both   longitudinal momentum fraction $x$ and  transverse 
 momentum k$_{\rm{T}}$ fixed~\cite{hef}.  
This  factorization program is carried through  in~\cite{jhep09}  
for   the case of  forward    jet hadroproduction.  
Applications of this  framework    
to forward  physics at the LHC  
are discussed in the contribution~\cite{deak}. 

    The factorization     in     $x$  and k$_{\rm{T}}$    in 
Fig.~\ref{fig:ypt}     
  is valid to single-logarithmic accuracy~\cite{hef}. 
In particular, 
it is consistent with the all-order factorization of collinear singularities~\cite{css-collfac}, 
allowing one to  fully control   the  dependence on the factorization scheme and scale.   
Conversely,  it enables   one to  obtain   logarithmically 
enhanced  terms in rapidity  that are   not associated to 
any collinear logarithm.  
This in contrast with  calculations  in  double-logarithmic approximations. 

\begin{figure}[t!]
\vspace{4.5cm}
  \begin{picture}(30,0)
    \put(30, -40){
      \includegraphics{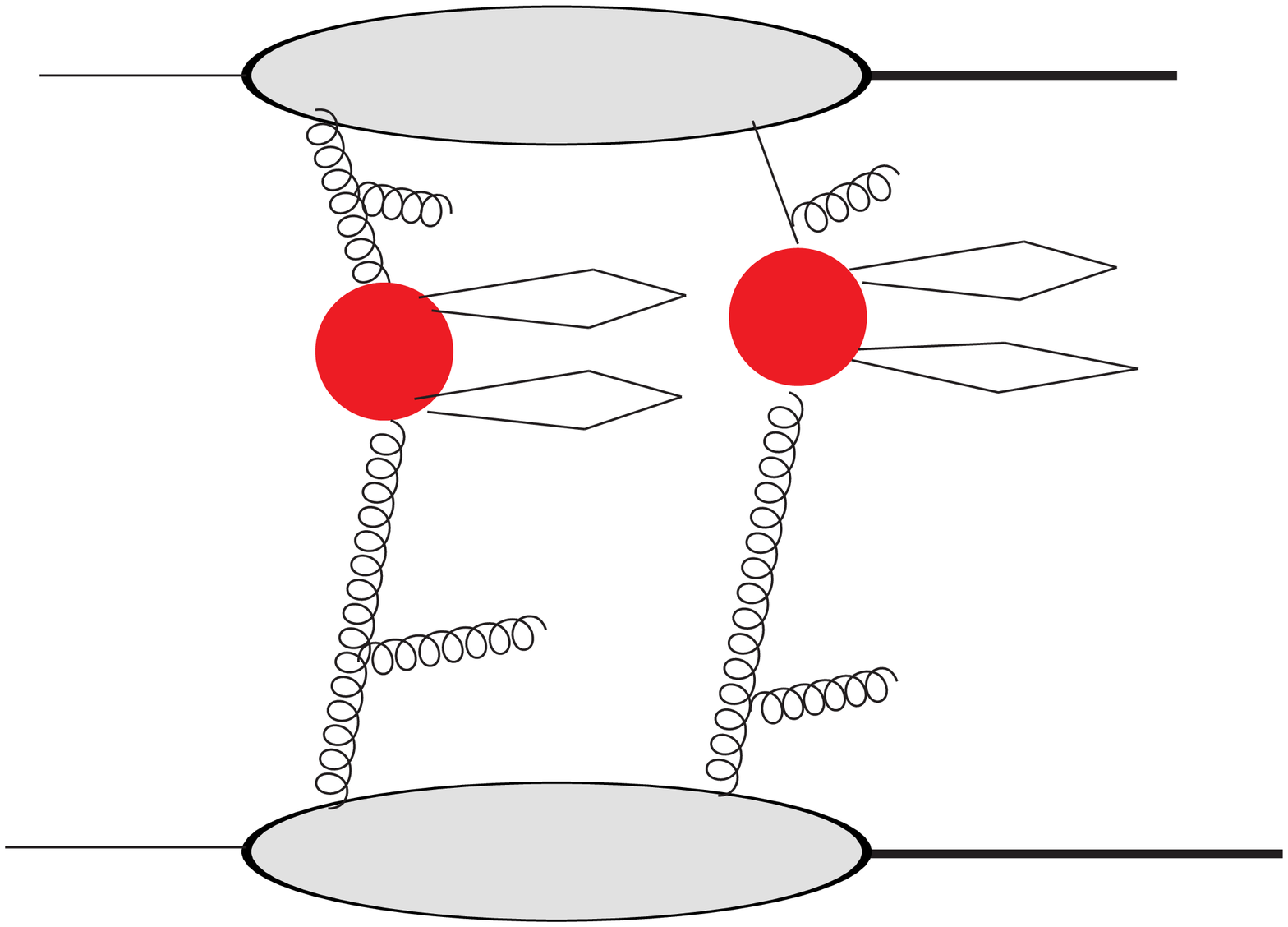}
    }
    \put(220, -40){
      \includegraphics{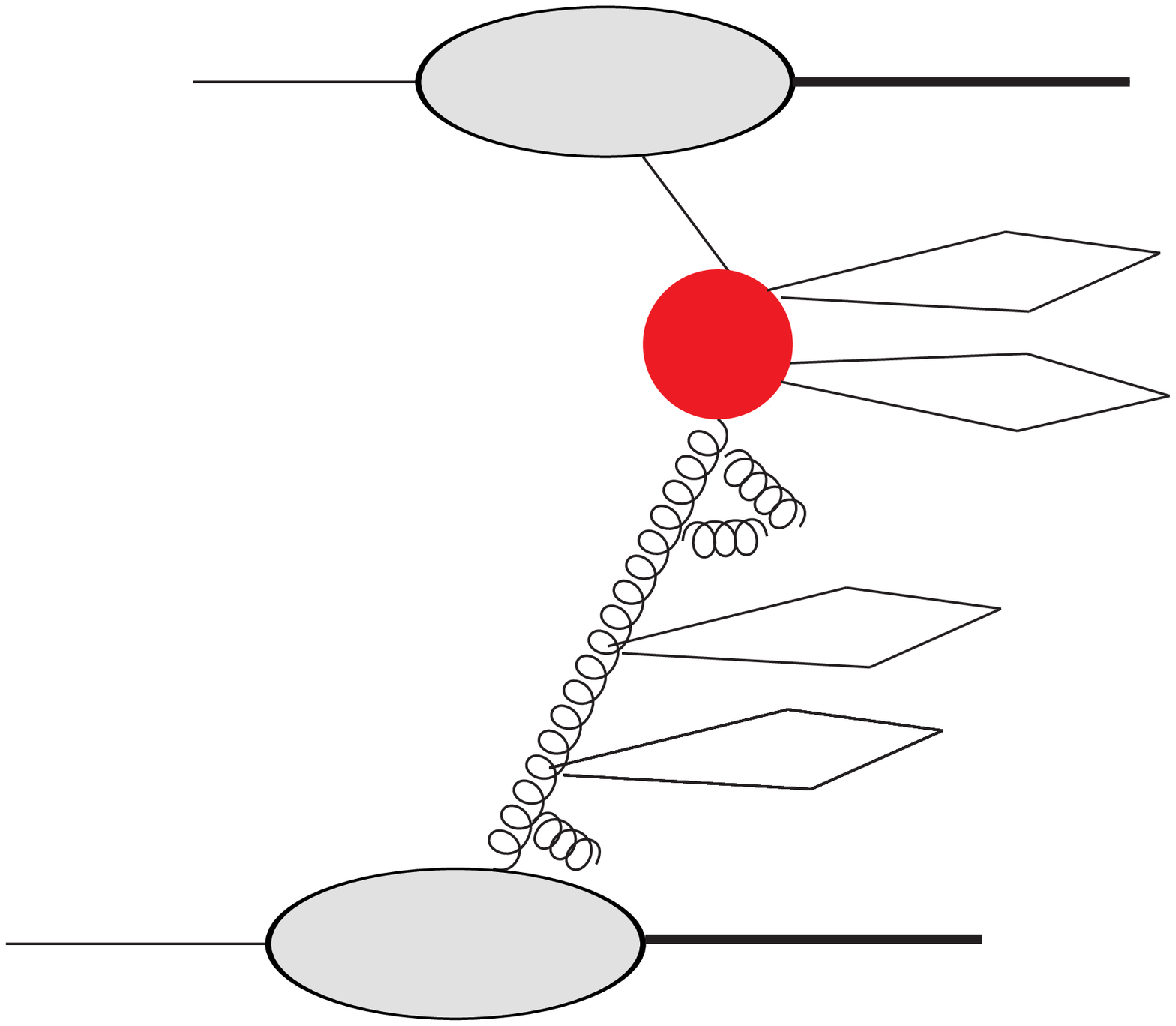}
    }    

     \end{picture}
\caption{\it  Multi-jet production by 
 (left) multiple  parton chains; (right) single parton chain.} 
\label{fig:mpi}      
\end{figure}

Besides the  different  classes of  all-order    radiative  corrections to 
single parton scattering illustrated in   Fig.~\ref{fig:ypt},   and the 
corresponding     perturbative QCD resummation methods, 
another  type of   dynamical  effects   which   it  has been argued may  have 
non-negligible impact  on forward  processes  involves contributions 
from multiple parton interactions~\cite{bartfano}, and is depicted in Fig.~\ref{fig:mpi}.  
This   picture   illustrates that   
the production of  final   states with multiple jets  may occur   
by interactions from a  
 single parton chain  or by interactions  from multiple chains.   
 The multi-parton interactions are  essential if one is to describe 
 minimum-bias   collider processes;   but they may also affect 
 significantly events  involving  a hard trigger~\cite{bartfano,gosta}.  
   Multi-parton  
 interactions are   modeled in the parton-shower   event generators 
used to simulate  final states  at the 
LHC~\cite{bartfano,gosta}, and are  the subject of 
a number of  current efforts~\cite{blok}   to    construct approaches capable of 
 incorporating  multiple scatterings in a  partonic  framework. 
  
A   basic question for    phenomenology    is  to what extent  
   current Monte Carlo generators    can   provide 
 realistic  event  simulations     of  forward particle production. 
Detailed   measurements  of forward-region observables, such as  
   those   discussed   in Sec.~\ref{sec:lhc},    
 should    enable one to make comparative studies of the  
  different  mechanisms  in   Fig.~\ref{fig:mpi}   for multi-jet production and   
  investigate   whether   QCD   effects   are 
  well described by current Monte Carlo tools. 
    Examples of such studies are  given  in Sec.~\ref{sec:further}.

As noted  earlier,    the forward  region   implies asymmetric  parton 
kinematics and      is sensitive  to  the structure of the  
 initial state    near   small   $x$.   
 Measurements of  
forward    high-p$_{\rm{T}}$   production  processes  
 could    in principle be used 
for   PDF  determinations~\cite{anderson}.   
As they probe  the gluon density function   for  small $x$,  
  they could 
    naturally be used  also 
 to investigate  possible   
 nonlinear  effects~\cite{marquet-talk,ianetal} 
 at  high parton density.   
 The formulation~\cite{jhep09}  of forward jets, 
  based on the factorization in both $x$ and k$_\perp$ in 
 Fig.~\ref{fig:ypt}, although it cannot   by itself  be used in the   high-density  
  saturation  region,      
 is however  well-suited for  describing  the approach to this   region,  since  
 it is designed to take  
  into account  both the  effects from  high-energy  (BFKL)  
    evolution associated with the 
  increase in rapidity  and  also  the 
 effects from   increasing    p$_{\rm{T}}$   
 described by  renormalization group, which are  found to be also quantitatively 
 significant~\cite{kov10}    for studies of  parton saturation.  
 First    Monte Carlo  calculations   along  these   lines, 
   attempting to include saturation effects,  are given in~\cite{absorpt}.   
   An important role   will be played by 
studies of  the forward region in 
     collisions of dense  systems. 
See~\cite{yuan11} for   recent work on   high-density effects in 
jet production   in proton-nucleus collisions.   

Finally, we note  that 
many of the   theoretical  issues  
   that underlie   forward  physics,   from     
  QCD resummations  to 
   parton showering  beyond leading  order   to  potential   effects of  
     parton saturation,       
 depend on   the notion of transverse momentum dependent, or unintegrated,    
parton distribution functions.   See   
contribution~\cite{igor} at this meeting  and~\cite{hj_rec} for recent overviews. 
Transverse momentum dependent  distributions are 
     currently at the center   of  much activity,   see    e.g.~\cite{becher-neub}, and     
   their  uses   cover a broad range of   QCD applications. Here 
  we limit ourselves to observing that,  in the context of forward  physics,   
formulations  at   unintegrated    level  may  
 provide  a    natural framework   if we are to   
 extend   the theory towards the soft,    low-p$_{\rm{T}}$ region~\cite{gosta}, and 
 treat  phenomena such as     diffraction   and the physics of 
 multiple gluon rescattering.    An example is the   
 discussion~\cite{enberg,enberg_prd}  of   hard-diffractive processes. 
Another example is the study~\cite{yuan11}  of  unintegrated PDFs for 
scattering on  dense targets. Also  note that 
techniques  are being developed~\cite{s-channel}    
to  incorporate  the treatment of     multiple-gluon rescattering graphs  at small $x$   
in the operator   matrix-element formalism~\cite{css-collfac,cs81-82}  for  
parton distribution functions. 
We finally recall that  unintegrated  parton  distributions are a building block in 
QCD models    for  central exclusive    production processes~\cite{watt_09}, 
which will complement, at later stages of the     LHC program when 
near-beam proton taggers are installed~\cite{albrow_review},    
   studies of  forward     high-p$_{\rm{T}}$  production  such as those discussed 
 above.

\section{First LHC measurements}   
\label{sec:lhc}

  The earliest   LHC  runs   yielded  first results on forward physics.  This section 
  describes  a   sample of these early results.   

\begin{figure}[htbp]
\vspace{65mm}
\includegraphics{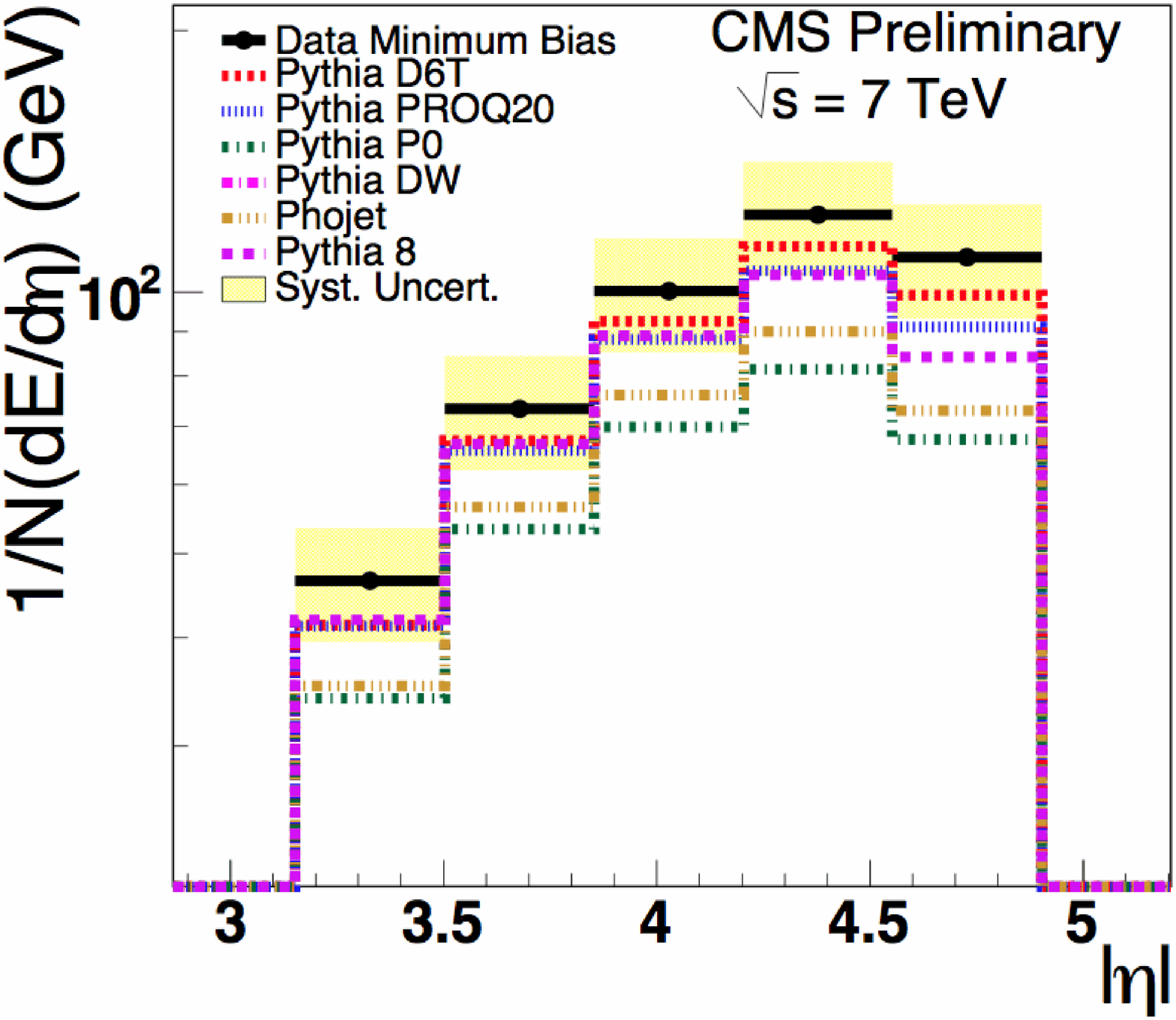}
\includegraphics{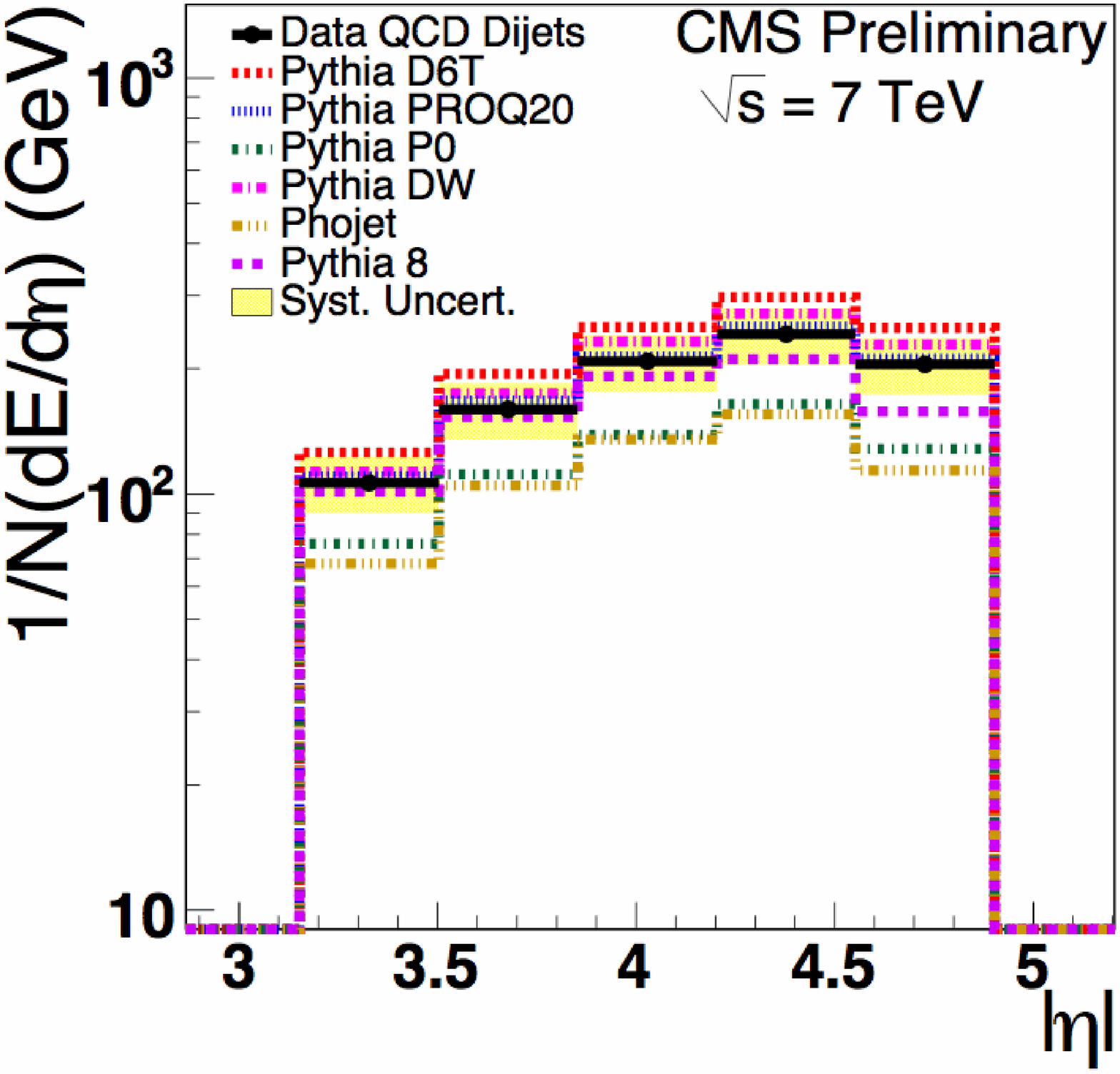}
\caption{\it    Energy flow in the minimum bias sample (left) and in the 
central dijet  sample (right)   as a function of pseudorapidity~\cite{sunar}. } 
\label{fig:sunarfigs}
\end{figure}

The CMS  Collaboration  reported  
   measurements of  forward energy and particle 
flow~\cite{sunar,bartalini-ichep,cms-pas-10-02}  for  pseudorapidity 
$ 3.15 < | \eta | < 4.9 $ at three 
different center-of-mass energies $\sqrt{s}$ of 
  0.9 TeV,   2.36 TeV  and    7   TeV.  The energy flow in the forward region 
 is measured  for minimum bias events and for events with 
 central ($| \eta | <  2.5 $)  dijets.   
 Fig.~\ref{fig:sunarfigs}~\cite{sunar}   shows results for $\sqrt{s} = $ 7  TeV.

The   energy flow is  observed to increase with increasing $\sqrt{s}$.  
  It is  found~\cite{sunar,cms-pas-10-02}   
that the observed   energy flow   in the forward region 
 is not  well  described by   tunes of  the \pythia\  Monte Carlo generator  based on 
  charged particle spectra in the central region, especially for the minimum 
bias sample.  

\begin{figure}[htbp]
\vspace{60mm}
\includegraphics{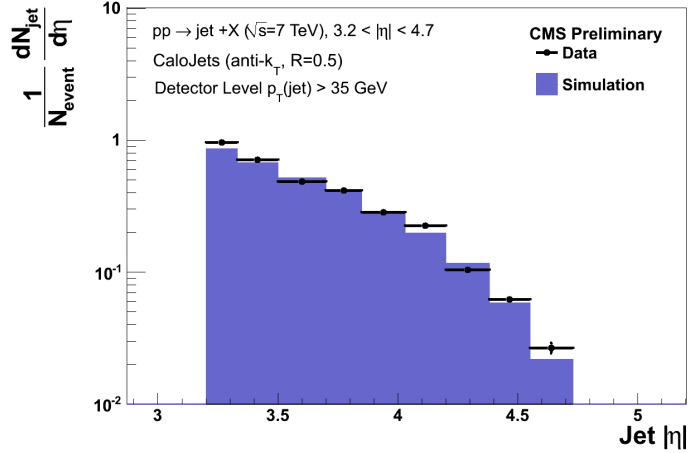}
\caption{\it    Detector level  forward jet   pseudorapidity  spectrum~\cite{sunar}. } 
\label{fig:sunar-jetrec}
\end{figure}

CMS also  reported first results on  reconstruction  of jets at 
forward rapidities~\cite{sunar,cms-dps-10-026}, 
with  35 GeV  $<$   p$_{\rm{T}}$  $<$  120 GeV   and  
  $ 3.2 < | \eta | < 4.7 $.   
  This is the first time that   jets are   observed  in hadron-hadron collisions 
  at such forward rapidities $\eta > 3$.   
    Fig.~\ref{fig:sunar-jetrec}~\cite{sunar} shows the detector level forward jet 
pseudorapidity  spectrum.

\begin{figure}[htbp]
\vspace{64mm}
\includegraphics{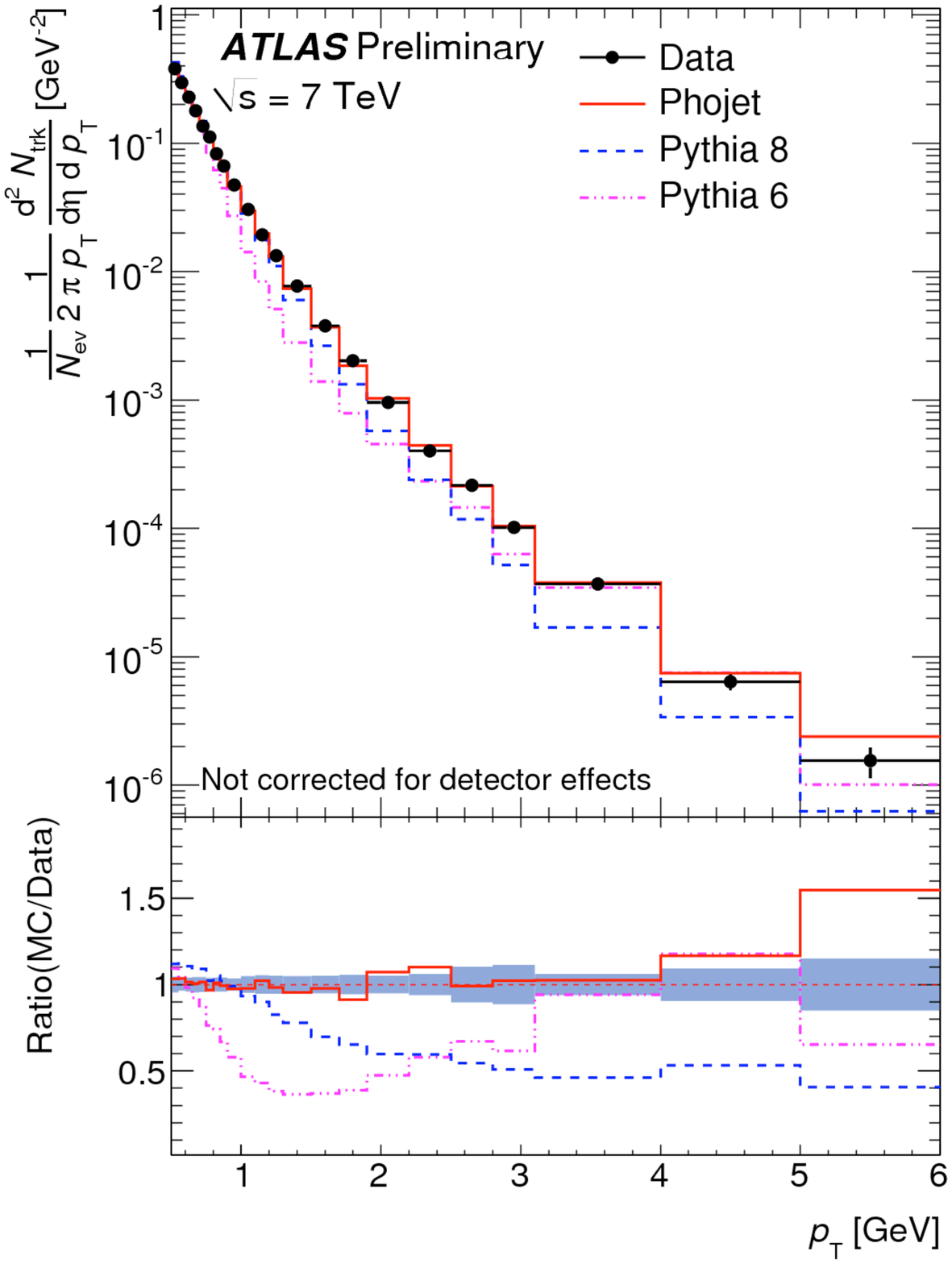}
\includegraphics{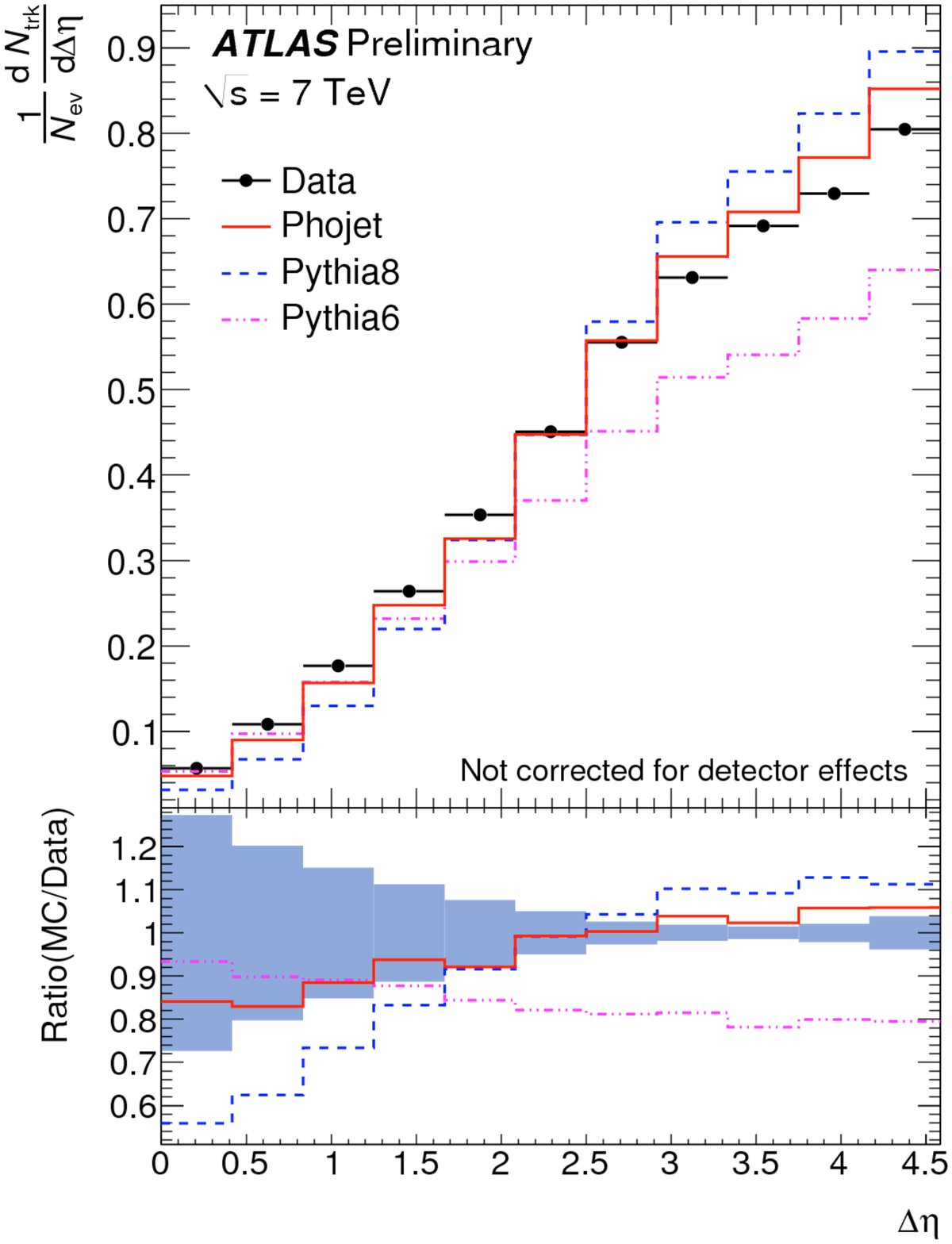}
\caption{\it  Track transverse momentum and rapidity gap distributions for   
a  diffraction enhanced   sample in minimum bias events~\cite{navin}. } 
\label{fig:navinfigs}
\end{figure}

The absence of energy deposition  
  in the forward region was used by 
ALICE, ATLAS and CMS to   identify 
diffractive events~\cite{navin,atlas-048,cms-001}.   
  Fig.~\ref{fig:navinfigs}~\cite{navin}  shows the ATLAS 
track   distributions   
in transverse momentum   p$_{\rm{T}}$  and rapidity gap 
 $ \Delta \eta $   for   the 
diffractive  fraction  in   minimum bias events~\cite{atlas-048,nurse-ichep}. 
Prospects for diffraction in  LHCb are discussed in~\cite{schmelling}. 

\begin{figure}[htbp]
\vspace{65mm}
\includegraphics{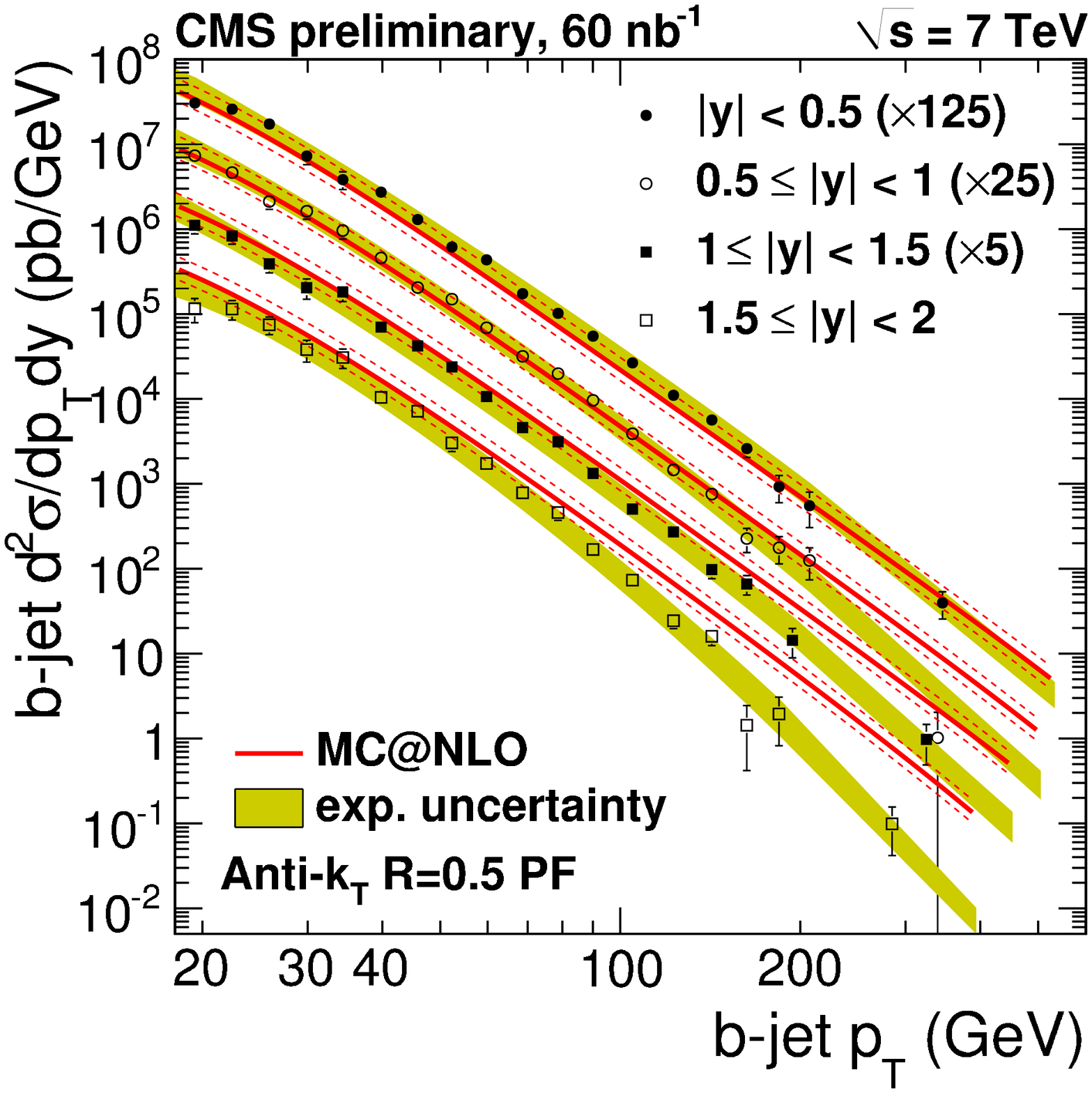}
\includegraphics{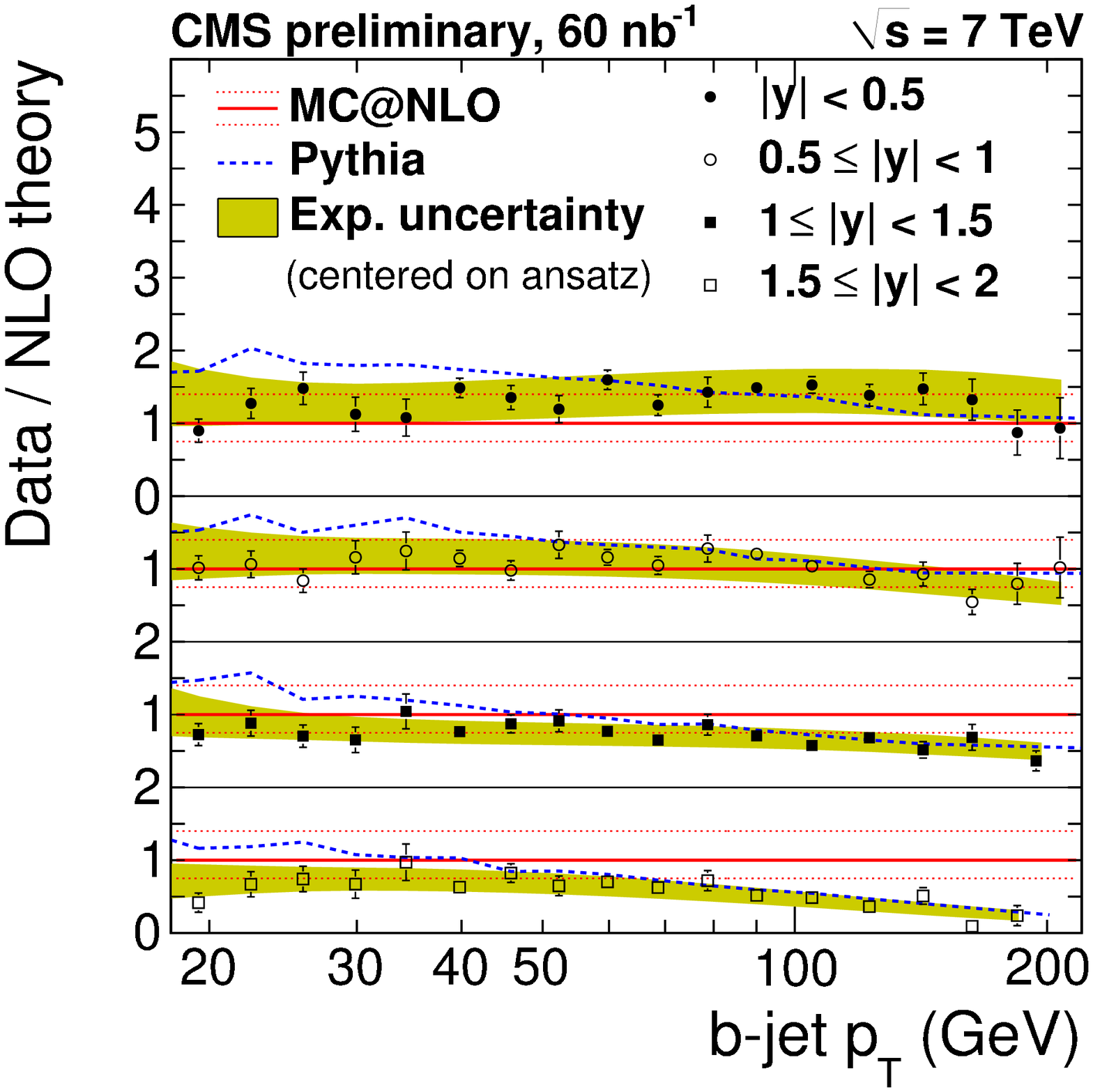}
\caption{\it   CMS  b-jet cross section versus p$_{{T}}$~\cite{cms-bph-009}. } 
\label{fig:cms-bjet}
\end{figure}

The CMS Collaboration performed measurements of $b$-jet  cross  
sections~\cite{cms-bph-009,cms-b-1101}. The  comparison 
 with  results of  the 
MC@NLO Monte Carlo  event generator  
in Fig.~\ref{fig:cms-bjet}~\cite{cms-bph-009} 
indicates  
potentially  interesting effects in the  $b$-jet 
transverse momentum   
spectra for the most forward rapidity bins. Reconstruction of $b$-jets at forward 
rapidities   will  also be  feasible  in LHCb~\cite{bay-potterat}. 
 These   studies  will  impact  
measurements of the Higgs to $ b {\bar b}$ decay channel in Higgs production 
associated with vector bosons.

\section{High energy cosmic rays}
\label{sec:cr}

In this section we  consider    the impact    of collider measurements 
on investigations
 of high-energy cosmic rays (CR). 
 
\begin{figure}[htb]
\vspace{85mm}
\includegraphics{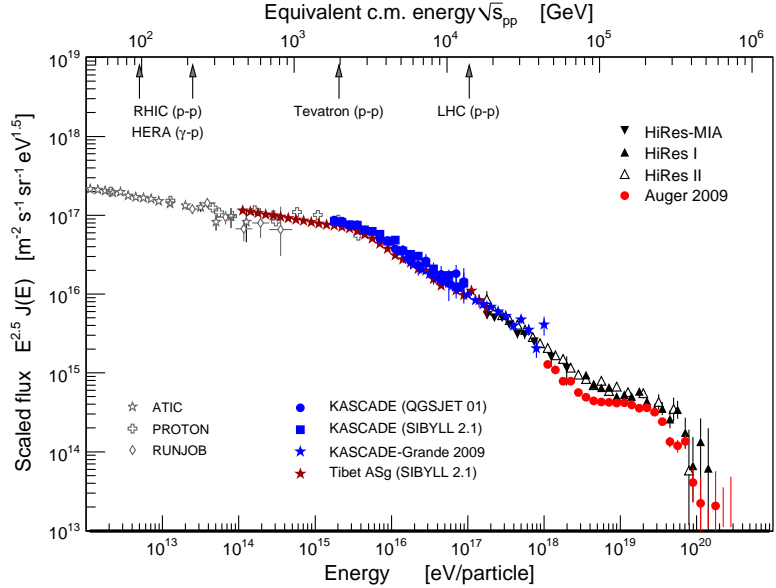}  
\caption{\it   The observed flux of high energy cosmic rays (from
Ref.~\cite{denterria11}).} 
\label{fig:CR-spec}
\end{figure}

 The high-energy CR spectrum has been measured  
 to laboratory energies in excess of   
 $10^{11}$ GeV~\cite{watson,bluemer,rodriguez10}.       
 The observed CR spectrum  is 
 shown in Fig.~\ref{fig:CR-spec}.   
 However, winning
 in the maximal energy,   Nature cannot compete with human-made accelerators   
 regarding the luminosity, which is especially true for   
  ultra-high energy cosmic rays (UHECR) at $E>10^{10}$ GeV, whose    
 flux is of the order of a particle per km$^2$ per century.    
 Hence, one is forced to use the atmosphere of the Earth as the target,   
 inferring the properties of the primary CR particles from the characteristics    
 of huge nuclear-electromagnetic cascades   (Fig.~\ref{fig:airsho}) 
   - extensive air showers (EAS),    
 induced by them in the air \cite{engel,watson,knapp03}.    
 Primarily one is interested in the CR arrival directions, the energy spectrum,
 and the elemental composition   (protons, nuclei, possible
 admixtures of photons and neutrinos).
 In fact, it is the latter which allows one to choose
 between various models for the origin of high and ultrahigh energy cosmic rays,
   providing in particular a decisive discrimination between astrophysical
   explanations of UHECR and scenarios which involve physics
   beyond the Standard Model \cite{sigl,kachel}.

 The classical air shower technique is to study the products of air shower 
 development in ground-based detectors, as performed, for example, by the 
 KASCADE  experiment \cite{kascade}. The most important observables in such a 
 case are densities of all charged particles (mainly electrons and positrons)
  and of muons  at the ground level. While the energy of the primary cosmic
  ray is typically reconstructed from the charged particle signal, the nuclear
  mass of the primary is inferred from the relative strength of the measured
  muon component, based on a higher multiplicity of charged hadrons, hence,
  of muons produced by pion and kaon decays, in nucleus-air collisions, compared
  to the case of the primary proton. Alternatively, one can study the longitudinal
  EAS development detecting the fluorescence light emitted by air molecules
  which are excited by charged particles propagating through the atmosphere,
  the corresponding technique introduced by the Fly's Eye experiment \cite{fly}.
 In such a case, the primary CR energy can be rather precisely deduced from
 the estimated total amount of fluorescence light while the type of the primary
 particle influences strongly the so-called shower maximum position  --
 the depth (in g/cm$^2$) in the atmosphere where the maximal number of
 ionizing particles is observed. Modern experiments, like the Pierre Auger
 Observatory \cite{PAO}, combine both techniques and use them to cross-calibrate
 the results. 
 
  Naturally, the so-obtained
 spectrum and, especially, the composition of the primary cosmic rays
  depend  crucially on the 
 validity of the cascade description by the corresponding Monte Carlo codes.
 The  least certain part of such simulation
 programs is the treatment of hadronic interactions, which has to be extrapolated
 over many orders of magnitude beyond the energies studied at 
 accelerators~\cite{engel,ulrichetal}.     
 Besides, as in any thick target experiment, particle densities at ground are
 most   influenced by the forward spectra of secondaries in hadron-air and
 nucleus-air collisions and by the  corresponding inelastic cross sections.
 Similarly, the longitudinal shower development is very sensitive to the
 magnitude of the inelastic proton-air cross section and to the very forward
 spectra of produced particles, in particular, to the relative fraction of
 diffractive collisions.
 
 Both the already obtained and the forthcoming LHC data have a great potential
 for improving EAS simulation procedures. In Ref.~\cite{denterria11}
 the predictions of hadronic Monte Carlo generators used in  the  CR field
 have been compared to the results of the CMS \cite{cms10} and ALICE \cite{alice10}
   collaborations
 on soft multiparticle production. While none of the models considered
 provided a sufficiently good description of the complete set of the available
  LHC data,
 the experimental results on the pseudorapidity density of produced charged hadrons
 appeared to be well bracketed  by the CR model predictions. As discussed 
 in Ref.~\cite{denterria11}, the experimentally
 observed smooth energy behavior of the multiplicity
 of charged hadrons  in the $\sqrt s =0.9\div 7$ TeV range    
  supports conventional astrophysical 
 interpretations of cosmic ray data in the discussed energy range,
  in particular   
 concerning    the  ``knee''  around
 $3\cdot 10^6$ GeV \cite{knee},  in contrast with 
 claims on exotic physics    (e.g.~\cite{petrukhin})      being responsible for the 
 observed features of the CR spectrum. 

 On the other hand, the LHC results   provide a firm ground for extrapolating
 existing hadronic interaction models to ultra-high energies.
 The accuracy of the description of air shower development will be further
 enhanced by the forthcoming LHC data. Especially important results
 are expected from the TOTEM experiment~\cite{totem} designed for high accuracy
 measurements of the total, inelastic, and diffractive proton-proton cross sections
 and from the studies of forward particle and energy flows by  the LHCf 
 experiment~\cite{adriani08,bonechi10}. The    LHCf experiment 
   concentrates on investigations of 
 forward spectra of neutrons and 
 photons~\cite{lhcf1012,lhcf-bongi}  and has a good potential
 to improve significantly model predictions for the spectra of leading
 baryons and mesons~\cite{lhcf-pi0}, which are presently extrapolated from fixed target
 energies. In turn, the good knowledge of the inelastic cross section and
 of forward hadron spectra in $pp$ collisions will significantly enhance the
 accuracy of the calculations of the longitudinal EAS development and 
 will offer an opportunity for precise studies of the CR composition by
 means of the fluorescence technique. 

As   LHC    forward production measurements  
 probe  the structure of the  initial state at   very  low  $x$, 
they  can potentially also  impact   predictions   for  the  scattering   of 
ultra-high energy neutrinos    
that are expected to accompany    UHECRs~\cite{quigg}.  
See~\cite{thorne1102,sarkars,berger} for recent evaluations of   
 structure function effects on  UHE  $\nu$  cross sections.   
 Given the size of the 
 theoretical uncertainties  estimated on the 
 neutrino cross sections~\cite{thorne1102} 
 from  the gluon  distribution, 
   it  can  be  relevant to exploit  both the   experimental information on low-$x$ 
   PDFs   that can be gleaned from LHC forward probes~\cite{anderson}   and 
the  theoretical constraints   on  the sea quark  distribution  
   associated with   multi-gluon rescattering   
 (see~\cite{hs-msbar} and first reference in~\cite{s-channel}).

\section{Further  collider studies}
\label{sec:further}

As  discussed in the previous  two sections,   
  the  LHC forward physics program
  is underway and is already providing useful results based on data collected in the earliest 
  machine  runs.    In this section we consider 
 further   aspects of this program         
   which can   be investigated     in the near future    based on the full 2010 data yield. 
  We do not address  here   issues    concerning   overlaid pile-up  events  
     from  the   increase in luminosity   in  2011 runs.  
 
\begin{figure}[htb]
\vspace{25mm}
\includegraphics{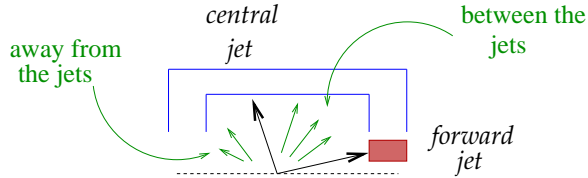}  
\caption{\it  Production of forward and central jets. } 
\label{fig:jetcorr} 
\end{figure}

 One   area  of investigation   concerns correlations of forward and central jets.   
The jet reconstruction capabilities of  forward + central detectors at the LHC 
give   the possibility to   study these   correlations  
in  rapidity,  azimuth    and transverse  momentum  (Fig.~\ref{fig:jetcorr}). 
These studies should   allow one to  
  probe    effects of  multi-gluon  radiation  
 across the large rapidity interval   and     
 in particular    make a comparative   investigation of  
   finite-angle, noncollinear   corrections  to parton showers   and  
  multi-parton interaction corrections~\cite{dhjk-prep}. 

 Measurements of forward-central jet correlations can be used for the QCD tuning of 
 Monte Carlo event generators. (For the counterpart of this in the case of central jet pairs 
 see the first LHC measurements~\cite{lhc-central-dijets}.)  
Specific  information on   jet distributions is to be gained  by going to 
 the forward region.  We observe that 
 this  will   be relevant also  in the case of heavy particle production, if 
 one is to use jets as a tool to  analyze potential effects of new physics 
 at the LHC from highly boosted massive states~\cite{boost}.

\begin{figure}[htbp]
\vspace{65mm}
\includegraphics{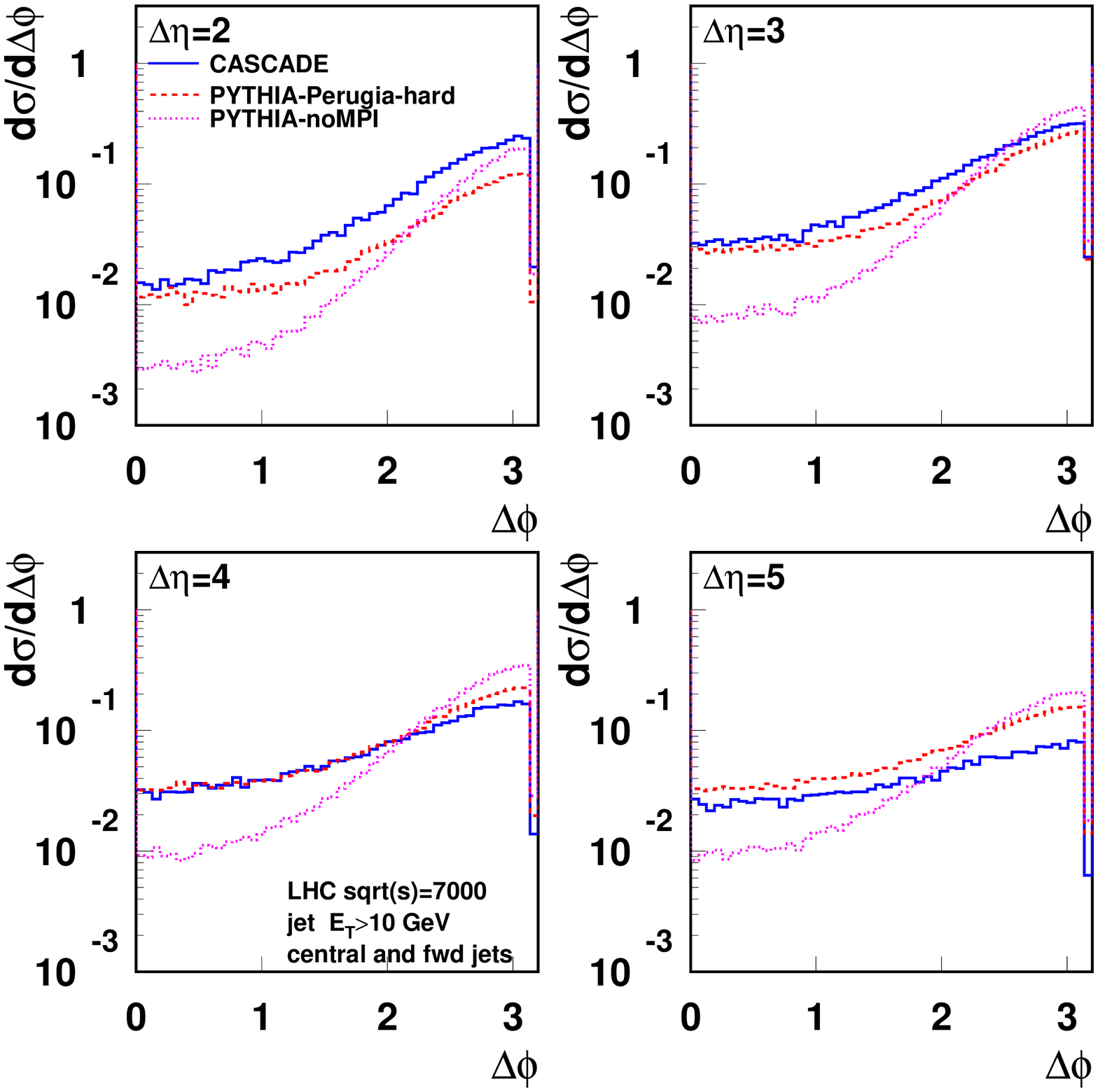}
\includegraphics{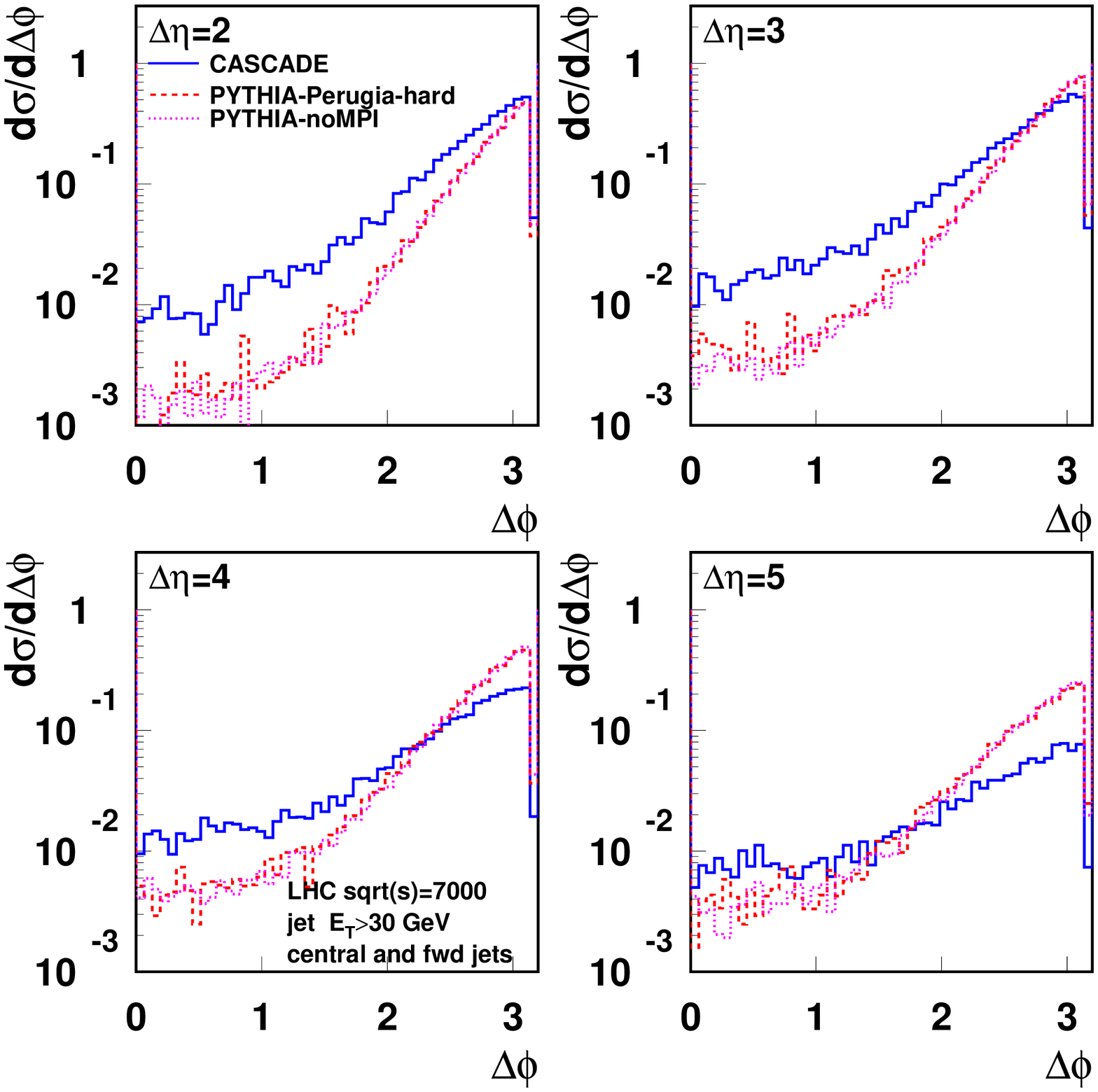}
\caption{\it Cross section versus  azimuthal separation 
$\Delta \phi$  between  central and  forward jet,  at  different  
rapidity  separations $\Delta \eta$,  for jets with transverse 
energy $E_T > 10$ GeV (left)  and 
$E_T > 30$ GeV (right)~\cite{dhjk-prep}. } 
\label{fig:azsigma}
\end{figure}

An example     is shown in 
Fig.~\ref{fig:azsigma} (from~\cite{dhjk-prep}),  where 
we see  the  cross section  
 as a function of the 
azimuthal separation $\Delta \phi$  between  central and  
forward jets reconstructed  with the      Siscone algorithm~\cite{fastjetpack} ($R =  0.4$)
  for different  rapidity separations. 
The solid blue  curve is the prediction based on   implementing  the factorization~\cite{jhep09}  
  in the  parton-shower event generator~\cite{cascadedocu} 
(\protect\CASCADE); the red and purple curves are  
 the predictions  based on calculations with 
  collinear  parton-showering~\cite{pz_perugia} (\protect\PYTHIA), respectively   
  including  multiple interactions  and without multiple interactions.  
 See comments in   Sec.~\ref{sec:th}  around   Fig.~\ref{fig:mpi}.    
     It is found in~\cite{dhjk-prep}   that 
 while the average 
 of the  azimuthal separation  
  $\Delta \phi$ 
 between the  jets is not  affected very  much 
   as a function of 
 rapidity  by   finite-angle gluon emissions, 
 the detailed shape of the  $\Delta \phi$ distribution is.    
 In particular 
 we see in  Fig.~\ref{fig:azsigma}  that 
the decorrelation as a function of 
$\Delta\eta$ increases in \CASCADE\ as well as in \PYTHIA , 
but  while  in the low $E_T$  region (Fig.~\ref{fig:azsigma} (left)) 
this   is similar  
 between \CASCADE\ and \PYTHIA\  with multiparton interactions   for $\Delta\eta < 4 $, 
in the higher $E_T$ region    
(Fig.~\ref{fig:azsigma} (right))
 the influence of multiparton interactions in \PYTHIA\ is small and 
\CASCADE\ predicts everywhere  a larger decorrelation
 as a result of finite-angle gluon  radiation in  single-chain parton shower. 
 
It is worth noting  that   jets in the forward region  are relevant 
not only for  LHC physics   but also  in the case of 
leptoproduction~\cite{mueproc90c}   
for  the physics   program 
at  the proposed  future   lepton facilities~\cite{laycock} (LHeC, EIC)   and 
 for further  analyses of ep HERA data~\cite{heraforw,polifka}.     
It can  be of interest to examine forward + central jets, similarly 
 to what  is described 
above for the LHC case,   also  in leptoproduction;     
 due to the  phase space available for  multiple jet radiation, 
 such studies  are likely to  prove more  relevant  
 at a future high-energy lepton collider than at HERA.  
(For   related  discussions of  central jet leptoproduction see~\cite{hj_ang}.) 
A  further,    interesting  possibility is    to examine forward jets associated 
with diffractive DIS~\cite{polifka}.

Another set of   potentially  interesting studies  at the LHC   concerns   
measurements   of   
particle and energy flow      in the regions both between the jets 
and away from the jets    in  Fig.~\ref{fig:jetcorr}.  
 As noted in~\cite{deak},   the inter-jet and the outside flows  
  would allow one to   gain more insight   into   the 
single-chain and multiple-chain mechanisms  of  Fig.~\ref{fig:mpi}.    
Especially,   one may  investigate quantitatively to what extent  
the multiple-interaction  case shifts   a significant amount of gluon  radiation 
 to larger  values of $x$ in the initial-state decay chains, as a result of less energy being 
 available to each of the sequential parton chains~\cite{dhjk-prep}. 
See also    the analyses~\cite{bryan_eperp_10} of energy flow observables.

An extension of   the studies discussed above  will involve 
  forward and backward jets.   Here one can   look   for   
Mueller-Navelet effects~\cite{denterria,ajaltouni,muenav,wallon-etal-10}.   
Investigating  QCD radiation associated with 
  forward-backward jets will serve to analyze 
 backgrounds    in Higgs searches      from 
 vector boson fusion channels~\cite{vbf}.   
In particular,     one may be able to extract information  on  
  Higgs couplings   by     
   studying the dependence on a  central jet veto~\cite{pilk}.  
In this case too     finite-angle radiative contributions to single-chain  showers,  
extending across  the whole  rapidity range, 
   affect the underlying  jet activity accompanying the Higgs~\cite{deak_etal_higgs} 
and  may  give competing effects to multiple-parton interactions.  

 Note that,  
  besides the case of jet production,  also  for  minimum bias  processes 
measurements of forward-backward correlations may  provide useful  information on  
 event structure~\cite{skands11}  and   properties of multi-particle production.

%\section{Acknowledgments}

%\section{Bibliography}

% ****************************************************************************
% BIBLIOGRAPHY AREA
% ****************************************************************************

% please do not change the following line
\begin{footnotesize}

% please do not change the following line
\end{footnotesize}

% ****************************************************************************
% END OF BIBLIOGRAPHY AREA
% ****************************************************************************

\end{document}